\documentclass[%
superscriptaddress,
 amsmath,amssymb,
 aps,
prb,
twocolumn,
floatfix,
]{revtex4-2}
\usepackage{mhchem}
\usepackage{gensymb}
\usepackage{booktabs}
\usepackage{graphicx}
\usepackage{dcolumn}
\usepackage{bm}
\usepackage[dvipsnames]{xcolor}
\usepackage{float}

\begin{document}

\preprint{APS/123-QED}

\title{Structural evolution of the kagome superconductors 
$A$V$_3$Sb$_5$ ($A$ = K, Rb, and Cs) through charge density wave order}%

\author{Linus Kautzsch}
 \affiliation{Materials Department, University of California Santa Barbara, Santa Barbara, CA, 93106, United States}%
 
\author{Brenden R. Ortiz}
 \affiliation{Materials Department, University of California Santa Barbara, Santa Barbara, CA, 93106, United States}%
 
\author{Krishnanand Mallayya}
 \affiliation{Department of Physics, Cornell University, Ithaca, NY, 14853, United States}
 
 \author{Jayden Plumb}
 \affiliation{Materials Department, University of California Santa Barbara, Santa Barbara, CA, 93106, United States}%
 
  \author{Ganesh Pokharel}
 \affiliation{Materials Department, University of California Santa Barbara, Santa Barbara, CA, 93106, United States}%
 
\author{Jacob P. C. Ruff}
 \affiliation{CHESS, Cornell University, Ithaca, NY, 14853, United States}%
 
 \author{Zahirul Islam}
 \affiliation{Advanced Photon Source, Argonne National Laboratory, Lemont, IL, 60439, United States}%
 
 \author{Eun-Ah Kim}
 \affiliation{Department of Physics, Cornell University, Ithaca, NY, 14853, United States}
 
\author{Ram Seshadri}
 \affiliation{Materials Department, University of California Santa Barbara, Santa Barbara, CA, 93106, United States}%
 
\author{Stephen D. Wilson}
 \email{stephendwilson@ucsb.edu}
 \affiliation{Materials Department, University of California Santa Barbara, Santa Barbara, CA, 93106, United States}%

\date{\today}

\begin{abstract}
The kagome superconductors KV$_3$Sb$_5$, RbV$_3$Sb$_5$, and CsV$_3$Sb$_5$ are known to display charge 
density wave (CDW) order which impacts the topological characteristics of their electronic structure. 
Details of their structural ground states and how they evolve with temperature are revealed here using single crystal 
X-ray crystallographic refinements as a function of temperature, carried out with synchrotron radiation.
The compounds KV$_3$Sb$_5$ and RbV$_3$Sb$_5$ present 2$\times$2$\times$2 superstructures in the $Fmmm$ space group 
with a staggered tri-hexagonal deformation of vanadium layers. CsV$_3$Sb$_5$ displays more complex 
structural evolution, whose details have been unravelled by applying machine learning methods to the scattering data.
Upon cooling through the CDW transition, CsV$_3$Sb$_5$ displays a staged progression of ordering from 
a 2$\times$2$\times$1 supercell and a 2$\times$2$\times$2 supercell into a final 2$\times$2$\times$4 supercell 
that persists to $T$ = 11\,K and exhibits an average structure where vanadium layers display both 
tri-hexagonal and Star of David patterns of deformations. Diffraction from CsV$_3$Sb$_5$ under pulsed magnetic 
fields up to $\mu_0H$ = 28\,T suggest the real component of the CDW state is insensitive to external magnetic fields.
\end{abstract}

\maketitle

\section{Introduction}

The charge density wave (CDW) instability is central to many of the unconventional properties reported in the 
$A$V$_3$Sb$_5$ ($A=$ K, Rb, and Cs) family of kagome superconductors \cite{Ortiz19,Ortiz20,Ortiz21superconductivity}. 
Within the kagome lattice of these materials, the CDW is hypothesized to harbor both real and imaginary components within the resulting order parameter, with 
the former corresponding to a real space charge inhomogeneity and the latter corresponding to a ``chiral'' flux 
phase that breaks time reversal symmetry \cite{PhysRevB.104.035142}. The real component of the CDW results in a 
superstructure deformation of the crystal lattice. Precise details of the structural pattern of this deformation 
has been an active area of recent investigation. 

\begin{figure}
	\centering
	\includegraphics[width=0.5\textwidth]{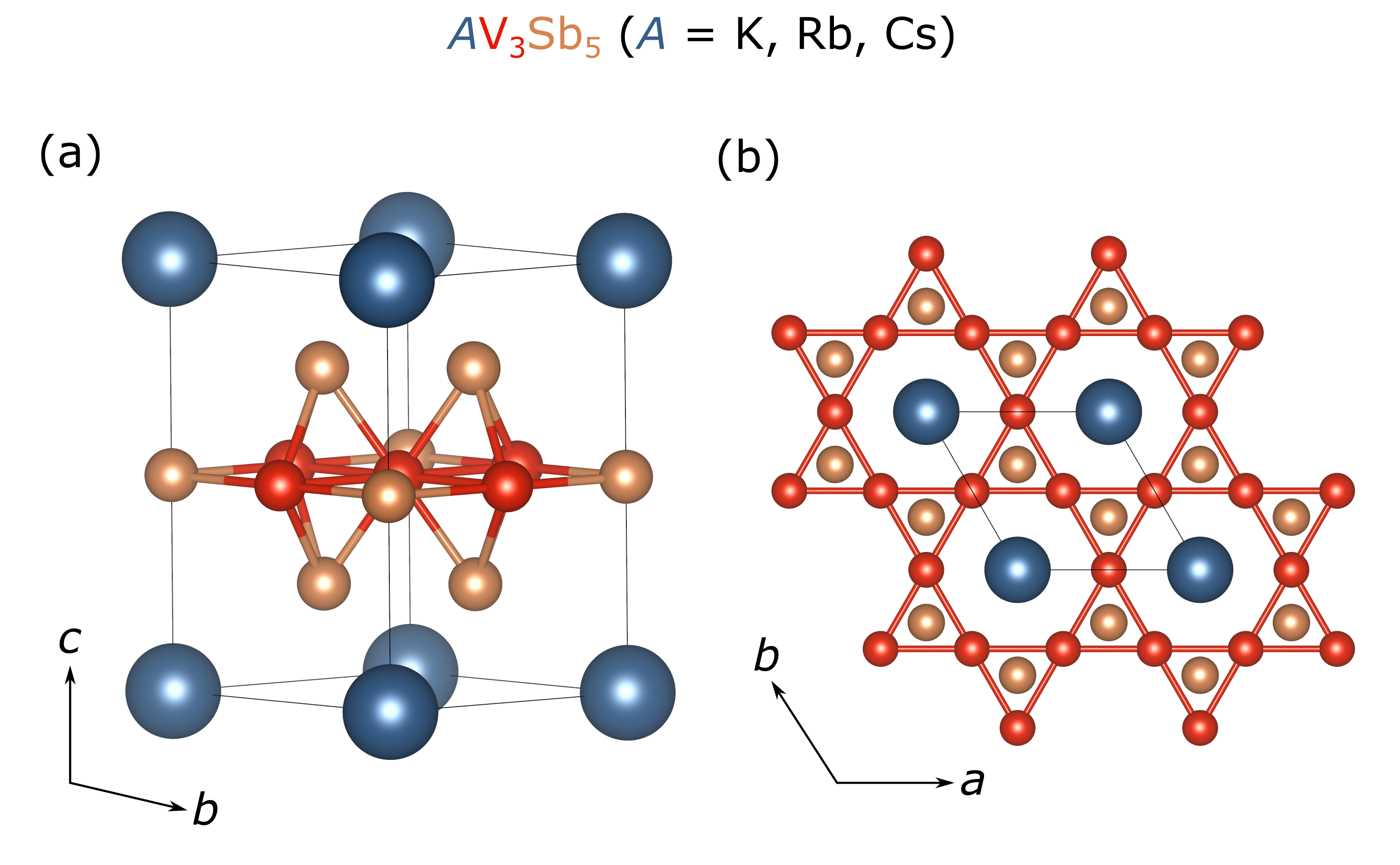}
	\caption{(a) and (b) show the refined room temperature ($T=290$~K) structure of the $A$V$_3$Sb$_5$ ($A=$ K, Rb, and Cs) family in the hexagonal space group $P6/mmm$.} 
	\label{fig1:rt_struc}
\end{figure}

Understanding the pattern of the real space structural deformation stands to inform a number of theoretical models of the anomalous CDW transtion in \textit{A}V$_3$Sb$_5$ compounds.  For instance, models of the band folding through the transition and calculations of the resulting Berry curvature may help inform the origin of the large anomalous Hall effect reported in the CDW state \cite{yang2020giant,yu2021concurrence}.  At high temperatures, the structure of all three variants assumes an ideal (undistorted) kagome network of vanadium ions in the space group $P6/mmm$ as shown in Fig.~\ref{fig1:rt_struc} \cite{Ortiz19}. The lattice deformation that results upon cooling through the CDW transition primarily arises from the displacement of the kagome network of vanadium atoms \cite{ortiz2021fermi}. The most energetically favored possibilities are predicted to be one of two breathing modes: either the Star of David-type (SoD) deformation or its inverse, a tri-hexagonal (TrH) deformation \cite{Jiang21,tan2021charge,christensen2021theory}.  Both are triple-\textbf{q} distortion modes, suggestive of the influence of nesting effects between the saddle-points ($M$-points) close to the Fermi level in \textit{A}V$_3$Sb$_5$. 

An added complexity in solving the pattern of CDW order in \textit{A}V$_3$Sb$_5$ is that interlayer interactions are strong enough to promote a fully three-dimensional CDW. This has been observed in scanning tunneling microscopy (STM) measurements as a half unit cell phase-shift between layers resolved across step-edges \cite{PhysRevX.11.031026} as well as direct observation of superlattice reflections with a finite $q_z$ propagation wave vector in x-ray diffraction measurements \cite{ortiz2021fermi,PhysRevX.11.031050}.  The out-of-plane $q_z$ periodicity varies between \textit{A}V$_3$Sb$_5$ variants with the unit cell doubled along the c-axis in \textit{A}=~K,~Rb variants \cite{Jiang21, PhysRevX.11.031050} and the unit cell at least partially quadrupled in \textit{A}=~Cs \cite{ortiz2021fermi, Stahl22, Xiao}.  The increased interlayer spacing driven by the larger Cs interstitial layer seemingly modifies the interactions between kagome planes and drives the formation of a delicate three-dimensional CDW phase whose modulation between kagome planes potentially hosts phase coexistence and metastable states.  

Metastability in the CDW phase of CsV$_3$Sb$_5$ has been primarily observed with x-ray diffraction measurements.  Initial reports found conflicting observations of $2\times2\times2$ \cite{PhysRevX.11.031050} and $2\times2\times4$ CDW superlattices \cite{ortiz2021fermi}, suggesting sensitivity to growth conditions.  Furthermore, recent x-ray diffraction studies have reported the coexistence of $2\times2\times2$ and $2\times2\times4$ structures with different onset temperatures and thermal quenching/annealing behaviors \cite{Stahl22, Xiao}.   Conclusions regarding which state is stabilized by annealing also vary, likely reflecting the added variable of disorder introduced via the growth process in the ground state selection.  For instance, recent studies have shown that small amounts of kagome-plane dopants can quench the $2\times2\times4$ state, leaving only a quasi-two-dimensional $2\times2\times2$ phase remaining \cite{kautzsch2022incommensurate}. 

The above complexity of the CDW state in CsV$_3$Sb$_5$ is relevant as it is the most easily grown and widely studied \textit{A}V$_3$Sb$_5$ variant. In our prior synchrotron study, an average structural solution for CsV$_3$Sb$_5$ was presented presuming a minimal model where the system maintains hexagonal symmetry ($P\bar{3}$). This average structure suggested a superstructure comprised of modulated TrH and SoD distortions in a $2\times2\times4$ cell \cite{ortiz2021fermi}. This prior solution could also be indexed as an orthorhombic cell; however twinning/domain formation was not directly resolved in the diffraction data. Recently, however, STM \cite{Li2022} and optical measurements \cite{Xu2022} have resolved three sets of orthorhombic twin domains in the CDW state of all three $A$V$_3$Sb$_5$ compounds.  This provides an excellent basis for developing twinned orthorhombic structural models of these systems via synchrotron x-ray diffraction measurements.

Here, we report the structural deformations resulting from CDW order in all three \textit{A}V$_3$Sb$_5$ compounds determined via single crystal synchrotron x-ray diffraction measurements. Structural refinements using an orthorhombic, three domain, twinned model are presented for all three compounds.  For KV$_3$Sb$_5$ and RbV$_3$Sb$_5$, a 2$\times$2$\times$2 superstructure is observed and structural refinement, combined with recent NMR analysis \cite{Sanna, Kun, Luo2022}, allows for a staggered TrH deformation of the vanadium layers to be resolved and the atomic positions determined. We also revisit the average structure of CsV$_3$Sb$_5$ in an orthorhombic cell using the same three domain structure, where we identify a 2$\times$2$\times$4 superlattice in the ground state.  The average low-temperature structural solution remains a modulation of SoD and TrH-type deformations, consistent with prior reports \cite{ortiz2021fermi, CominCVSRVSKVS, MingShiCVS}. Curiously, we also resolve a cross-over regime below $T=94$~K where a 2$\times$2$\times$1 and a 2$\times$2$\times$2 lattice form and then diminish at lower temperatures.  Finally, as a test for potential time reversal symmetry breaking in the charge density wave order parameter, we explore the field dependence of the superstructure in CsV$_3$Sb$_5$ under the application of a $\mu_0 H=$~28~T pulsed magnetic field.



\section{Methods}

\subsection{Structural solutions}

Synchrotron x-ray diffraction experiments were carried out at the QM2 beam line at CHESS. The incident x-ray wavelength of $\lambda = 0.41328$~$\rm \AA$ was selected using a double-bounce diamond monochromator. A stream of cold flowing helium gas was used to cool the sample. The diffraction experiment was conducted in transmission geometry using a 6-megapixel photon-counting pixel-array detector with a silicon sensor layer. Data was collected in full 360$^{\circ}$ sample rotations with a step size of 0.1$^{\circ}$. Scattering planes in reciprocal space were visualized using the NeXpy software package. The diffraction data was indexed and integrated using the APEX3 software package including absorption and extinction correction. Crystallographic structural solutions were determined using the SHELX software package \cite{shelx08}. An unsupervised machine learning tool using Gaussian Mixture Models called X-ray Temperature Clustering (X-TEC)~\cite{Venderley2022Proc.Natl.Acad.Sci.} was used to identify classes of peaks in reciprocal space with distinct temperature dependencies.

\subsection{Pulsed magnetic field experiments}

Pulsed magnetic field experiments were carried out at the beam line 6 ID-C of the Advanced Photon Source at Argonne National Laboratory using a dual cryostat, single solenoid pulsed-magnet system described in Ref.~\cite{Islam12}. X-rays with a wavelength of $\lambda=0.56356$~$\rm \AA$ were used, and the crystallographic $c$-axis of CsV$_3$Sb$_5$ was aligned parallel to the direction of the  incident x-ray beam and the direction of the magnetic field. A low-noise Lambda 750K detector was used with $55 \times 55$ $\mu$ m$^{2}$ CdTe pixels providing near 100\% efficiency. The detector was run in 12-bit mode at a 1 kHz repetition rate with XRD pattern collected every 1 ms. Multiple frames were collected over the course of repeated 9~ms pulses with a maximum magnetic field of $H_\mathrm{max}=28$~T. 

\begin{figure*}
	\centering
	\includegraphics[width=1\textwidth]{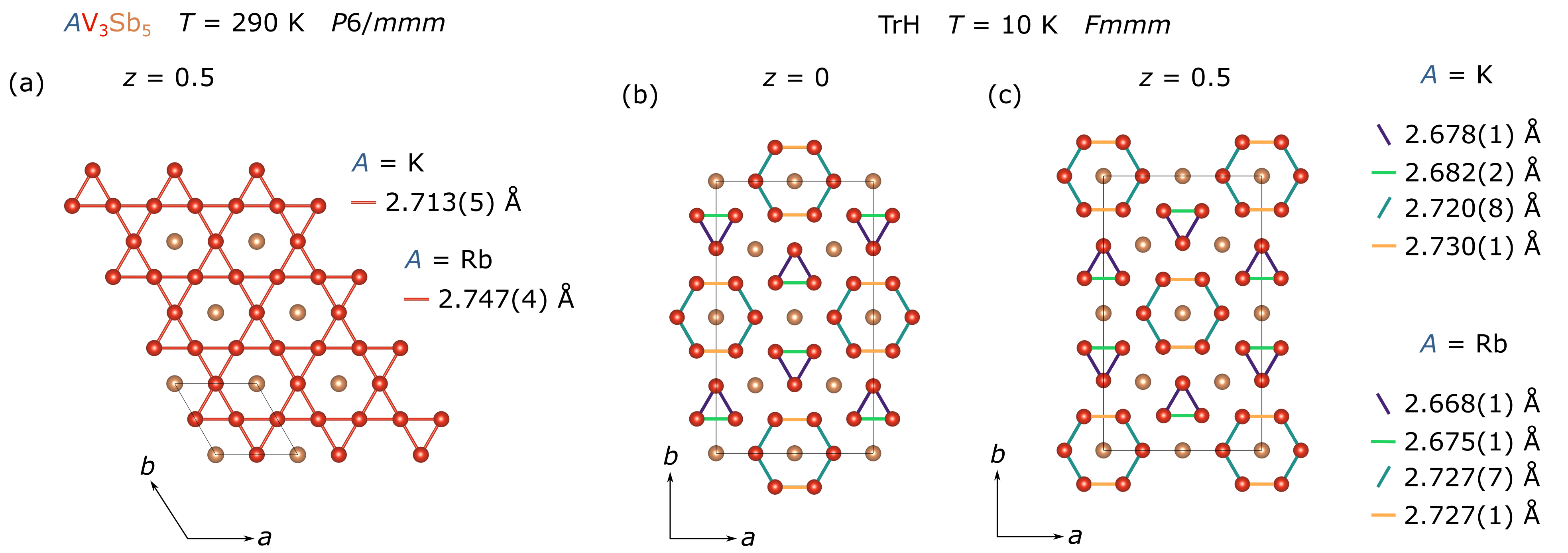}
	\caption{Refined structural models for KV$_3$Sb$_5$ and RbV$_3$Sb$_5$. (a) shows the z = 0.5 layer comprised of vanadium and antimony atoms of the refined room temperature structure of KV$_3$Sb$_5$ and RbV$_3$Sb$_5$ in the hexagonal space group $P6/mmm$. (b) and (c) show the suggested TrH model for the 2$\times$2$\times$2 superstructure in the orthogonal space group $Fmmm$. The vanadium planes at $z=0$ and $z=0.5$ in the $Fmmm$ structure are displayed. }
	\label{fig2:rvs_kvs_ref}
\end{figure*}

\section{Experimental Results}

\subsection{Structure of KV$_3$Sb$_5$ and RbV$_3$Sb$_5$}

We begin by discussing diffraction experiments on KV$_3$Sb$_5$ at $T=10$~K.  Scattering data indicate a 2$\times$2$\times$2 superstructure, and, similar to prior measurements \cite{ortiz2021fermi}, the area detector resolution does not allow for the momentum-space resolution of twinning in the distorted state.  Recent scanning optical studies have shown the presence of three structural twins that form below the CDW transition \cite{Xu2022}. These domains have their principle axes rotated by 120$^{\circ}$ in the ab-plane, consistent with an underlying orthorhombic lattice. This would imply a pseudohexagonal twinning with three orthorhombic twins that are rotated 120$^{\circ}$ with respect to each other in the $ab$-plane, which in turn raises the apparent symmetry of the diffraction pattern to hexagonal. Theoretical investigations identified the space groups $Cmcm$ \cite{subedi2022hexagonal} and $Fmmm$ \cite{christensen2021theory} as possible starting orthorhombic solutions to the twinned pseudohexagonal structure. 

As reflection intensities indicate a Laue class of $mmm$, model refinements were performed in both the face-centered orthorhombic space group $Fmmm$ and the base-centered space group $Cmmm$. Refinement in both space groups yields identical deformations of the vanadium layers with the distinction being that $Cmmm$ allows for unique distortions within neighboring vanadium layers.  Refinement in $Cmmm$ also implicitly allows for any intensity at positions forbidden in $Fmmm$ to be accounted for outside of twinning models.  Analysis in these two possible groups gave comparable or slightly better $R_1$ values for the higher symmetry $Fmmm$ space group, and this group was used for the final low-temperature structural determination.  

KV$_3$Sb$_5$ data refined in $Fmmm$ is well-represented by two structural models (Tab.~\ref{tab:refin}) consisting of either a staggered TrH ($R_1=6.22$, Fig.~\ref{fig2:rvs_kvs_ref}) or a staggered SoD distortion ($R_1=6.31$) of the vanadium layers. We note here that the assignments of TrH or SoD-type distortions in this paper are based on the dominant displacement type of V atoms, and that in all solutions there is detectable variance amongst V-V distances within a single plane (ie. there is not a single V-V distance in a given distorted kagome plane).  This is further parameterized in Discussion section of this manuscript. The pattern of a given distortion type is shifted by half of a lattice constant ($0.5a$) with respect to neighboring layers, and the two solutions differ in the sign of the displacement vectors of the vanadium atoms in the $ab$-plane. 

X-ray refinement alone is unable to discriminate between the distortion types, which is a result of the previously discussed twinning. However, recent NMR studies identify a TrH-type distortion and allow us to break the degeneracy between these solutions \cite{Sanna, Kun}.  The resulting low-temperature structure generated from the staggered TrH solution of KV$_3$Sb$_5$ is plotted in Fig.~\ref{fig2:rvs_kvs_ref} (b).  This is consistent with recent studies \cite{Stahl22,ratcliff2021coherent,tan2021charge}, and crystallographic information files (CIF) for both TrH and SoD distortion types can be found in the supplemental information. After collecting data at 10 K, the sample was warmed to 290 K and the high temperature structure measured.  This undistorted state was refined in $P6/mmm$ at 290 K with bond lengths illustrated in Fig.~\ref{fig2:rvs_kvs_ref} (a).

Similar analysis was employed for the x-ray diffraction data of RbV$_3$Sb$_5$, where RbV$_3$Sb$_5$ also forms a 2$\times$2$\times$2 superstructure in the CDW state.  At $T=10$~K, the data is again best represented when indexed in the space group $Fmmm$, and our refinements show superior solutions using a staggered TrH model ($R_1=6.40$) versus a staggered SoD model ($R_1=7.41$). This is consistent with NMR results \cite{Sanna}, and the TrH solution is selected as the correct structure for presentation and the resulting bond lengths in the vanadium network are shown in Fig.~\ref{fig2:rvs_kvs_ref} (b).  For reference, CIF files for both TrH and SoD distortion types can also be found in the supplemental information \cite{ESI}.  After measurement at 10 K, the 290 K structure was determined at 290 K in $P6/mmm$ with bond lengths illustrated in Fig.~\ref{fig2:rvs_kvs_ref} (a). 

\subsection{Structure of CsV$_3$Sb$_5$}

\begin{figure}
	\centering
	\includegraphics[width=0.5\textwidth]{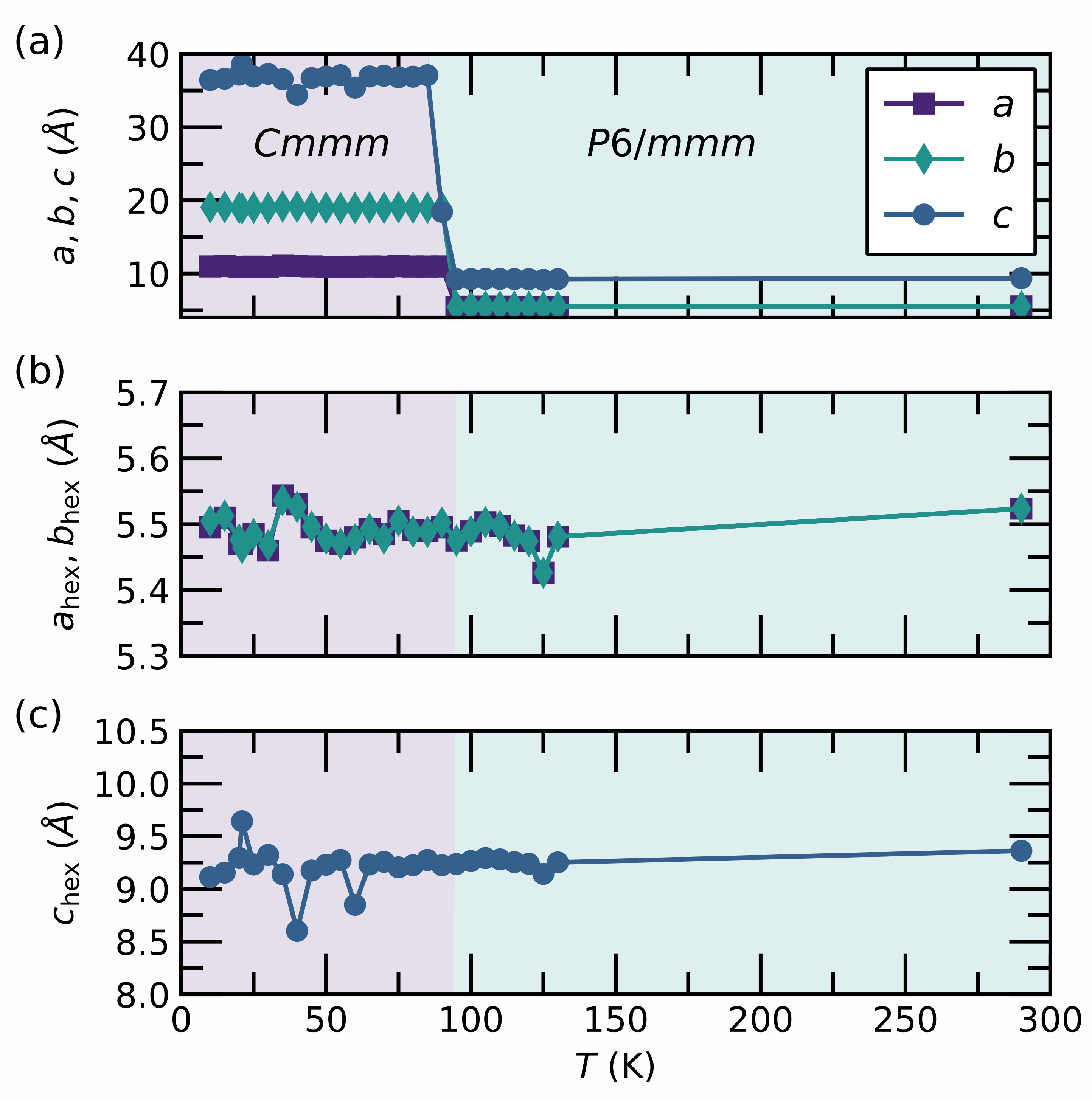}
	\caption{Temperature dependence of the unit cell parameters of CsV$_3$Sb$_5$ upon traversing the CDW transition.  The evolution of the $a$-, $b$-, and $c$-axes are plotted in panel (a). (b) shows $a$ and $b$ where $a_\mathrm{hex}=a/2$ for $T \leq 90$~K and $b_\mathrm{hex}=\sqrt{b^2 + 4a_{hex}^2}/4$ for $T \leq 90$~K and (c) shows $c$ where $c_\mathrm{hex}=c/2$ for $T = 90$~K and $c_\mathrm{hex}=c/4$ for $T < 90$~K. The subcell parameters undergo no significant changes.}
	\label{fig3:cvs_t_dep}
\end{figure}

\begin{figure}
	\centering
	\includegraphics[width=0.5\textwidth]{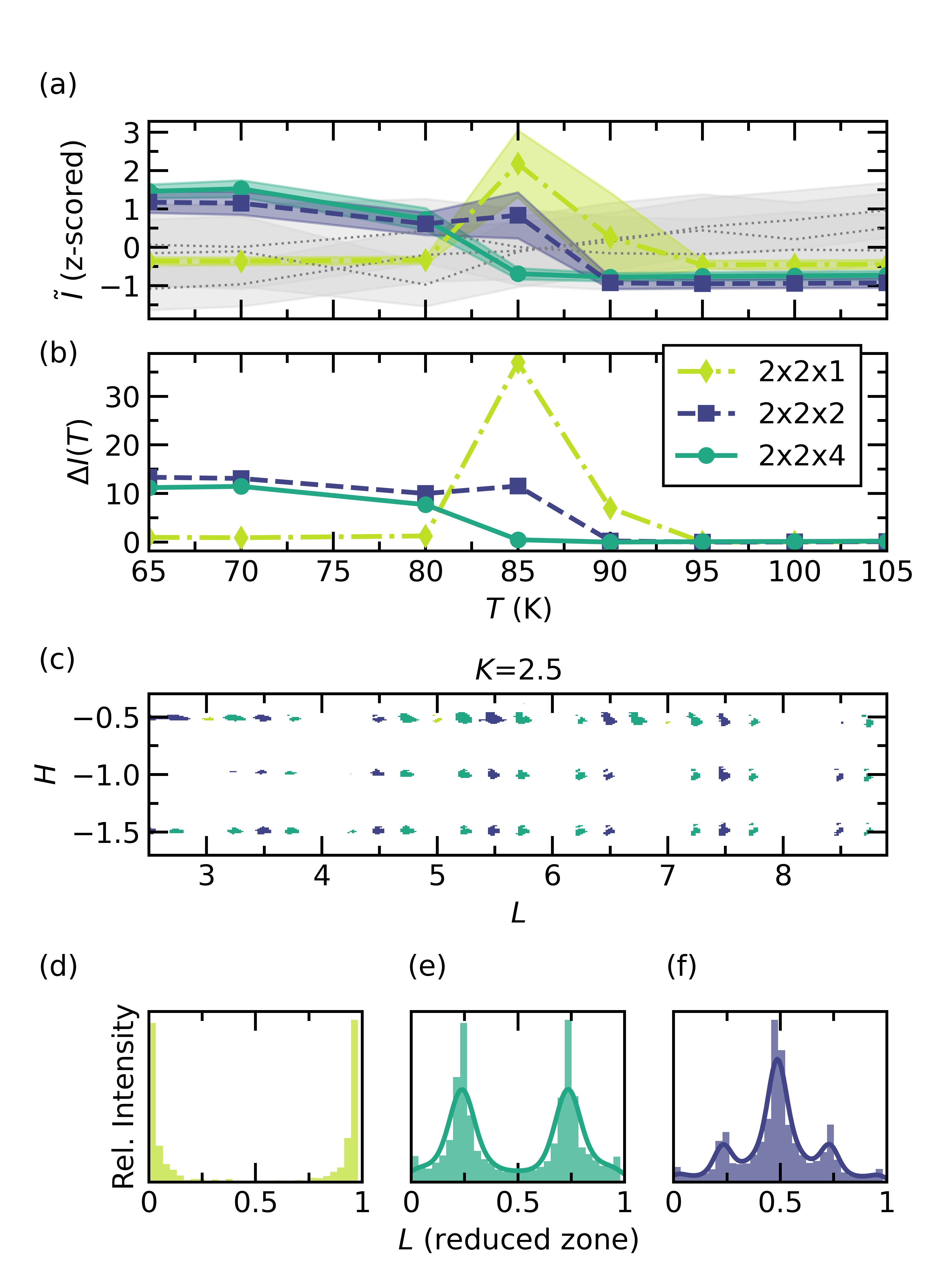}
	\caption{X-TEC analysis of CsV$_3$Sb$_5$ diffraction data. (a) X-TEC identifies three distinct temperature trajectories (blue, green, and yellow clusters) whose intensities exhibit a sharp onset near the CDW transition temperature. The lines denote the mean and shading denotes the standard deviation of the re-scaled (z-scored) intensities $\tilde{I}(T)$ within each cluster. The grey clusters identify the diffuse background scattering. (b) The average intensity of the blue, green, and yellow cluster reveals the different onset temperatures for the $2\times2\times4$ CDW ($T_{2\times2\times4}\approx 85K$) and $2\times2\times2$ CDW ($T_{2\times2\times2}\approx 90K$) state and the $2\times2\times1$ state emerging between 95~K and 80~K. $\Delta I(T)$ is the average intensity after subtracting its minimum value. (c) A section of the (H, 2.5, L) plane, in reciprocal lattice units (r.l.u.) and indexed in P6/mmm, showing the pixels colored by the cluster assignments of their intensities. (d,e,f) Distribution of the out-of-plane momentum $L$ of the peaks in the respective clusters.}
	\label{fig4:xtec_clustering}
\end{figure}

Now turning to analysis of the structure of CsV$_3$Sb$_5$, temperature dependent synchrotron x-ray diffraction experiments were carried out on a single crystal of CsV$_3$Sb$_5$.  The final data set was collected at $T=290$~K where data were indexed in the hexagonal space group $P6/mmm$. The Bragg peaks of the CsV$_3$Sb$_5$ crystal studied are narrow with minimal stacking disorder, though the crystallinity is anisotropic.  Quantifying this with Gaussian fits to representative Bragg peaks shows minimum correlation lengths in the $ab$-plane of $\xi_{min}=890$~$\rm \AA$ or $\xi_{min}=1000$~$\rm \AA$ (depending on the pixel-limited cut direction) and correlation lengths of $\xi=400$~$\rm \AA$ along the $c$-axis, indicating excellent crystallinity of the sample \cite{ESI}. The lower correlation length along the $c$-axis (interlayer direction) is typical in layered materials and beyond the resolution of experiments using relaxed collimation. 

\begin{figure}
	\centering
	\includegraphics[width=0.5\textwidth]{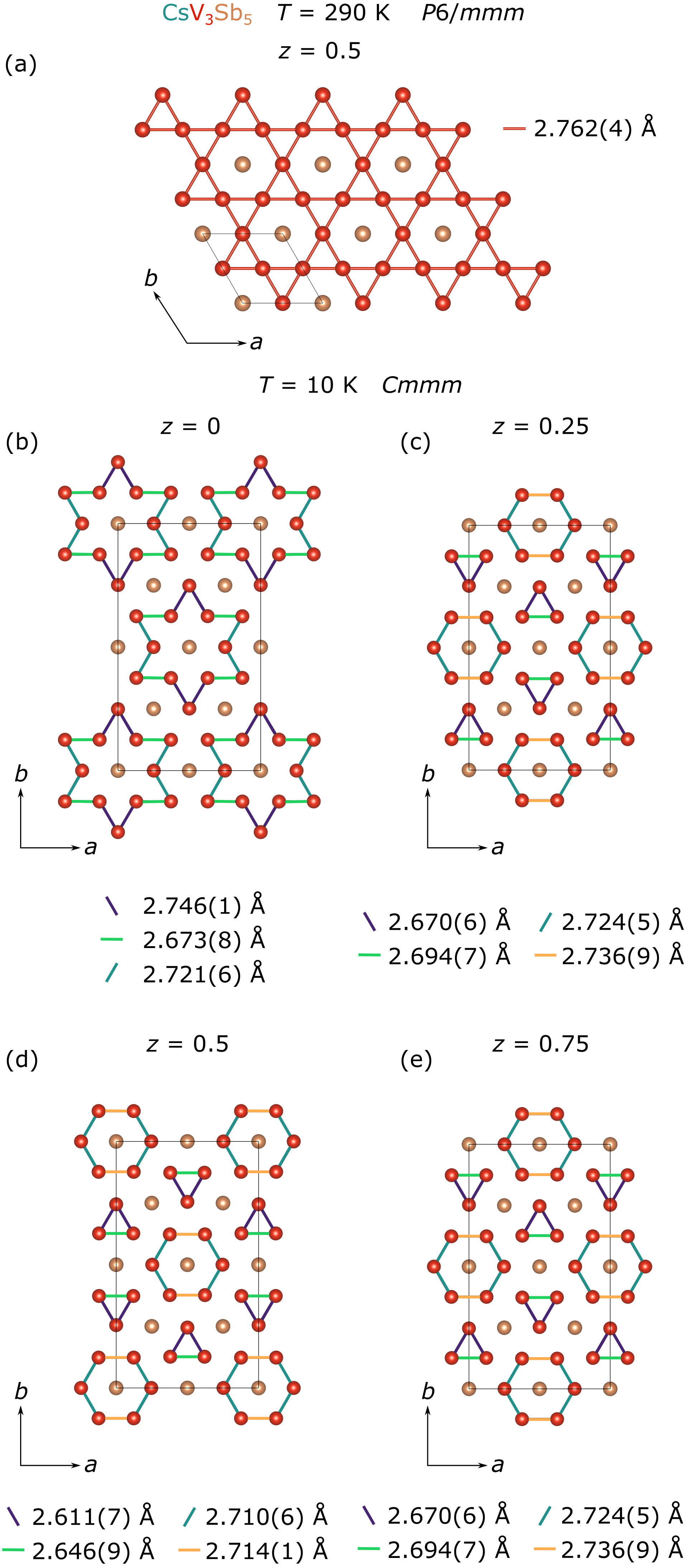}
	\caption{Refined structural model for CsV$_3$Sb$_5$. (a) shows the kagome network (z = 0.5) of the 290 K room temperature structure in the hexagonal space group $P6/mmm$. (b), (c), (d) and (e) represent a single-phase model of the 2$\times$2$\times$4 superstructure in the orthorhombic space group $Cmmm$. Three kagome layers form with TrH deformations that are capped by a layer with a predominant SoD deformation.}
	\label{fig5:cvs_ref}
\end{figure}

Temperature-dependent diffraction data were collected on warming after using an initial cooling rate of approximately 10~K/min to cool to 10 K. Data were then collected by warming at $\approx 5$ K/min with pauses for measurement scans (12~min/scan) performed at select temperatures. Collecting a data series using this cooling/warming profile show that at temperatures below $T=94$~K, superstructure reflections appear as expected. Just below the CDW transition, a clear staging behavior is observed.  At 90 K, superlattice reflections consistent with a 2$\times$2$\times$2 supercell are observed due to the appearance of half-integer $L$ peaks as well as half-integer peaks in the ($H$, $K$)-plane. Below this temperature, at 85 K, weak quarter-integer $L$ peaks also appear, signifying the onset of an enlarged 2$\times$2$\times$4 supercell.  The cell parameters through the CDW transition are plotted in Fig.~\ref{fig3:cvs_t_dep} 

\begin{figure}
	\centering
	\includegraphics[width=0.5\textwidth]{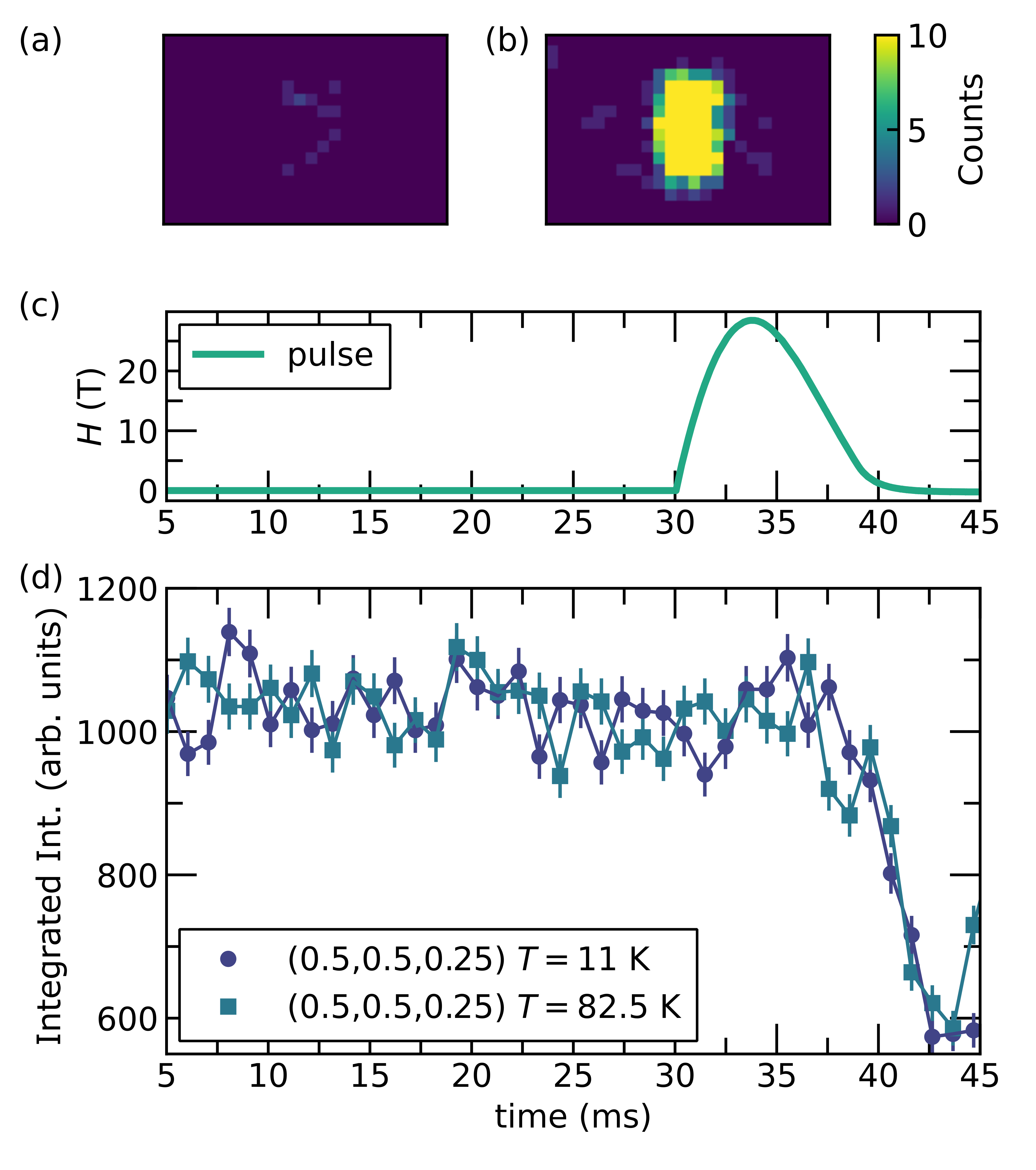}
	\caption{Magnetic field dependence of x-ray diffraction of the \textbf{Q}=(0.5,0.5,0.25) CDW superlattice reflection in CsV$_3$Sb$_5$ in a pulsed magnetic field experiment with $H_\mathrm{max} = 28$~T at $T = 11$~K and $T = 82.5$~K. (a) and (b) CDW reflection at $T=82.5$~K and $t=5$~ms from 1 sweep (1~ms exposure in (a)) and integrated over 61 sweeps (61~ms exposure in (b)). (c) Magnetic field $H$ as a function of the experiment time in milliseconds. (d) Sum of the integrated intensity of the (0.5,0.5,0.25) superlattice reflection from a total of 45 pulsed field sweeps at $T = 11$~K and 61 sweeps at $T = 82.5$~K.}
	\label{fig6:cvs_h_puls}
\end{figure}

It is clear that the nature of the charge density wave and the potential coexistence/competition between different states involves subtle interactions and changes in the diffraction patterns. In addition, the large temperature-dependent and \textbf{Q}-dependent data sets contain substantial quantities of information that can be overlooked during standard data reduction and integration processes. Recently, an unsupervised machine learning tool called X-ray Temperature Clustering (X-TEC) was developed to enable analysis of parametric, complex data sets. X-TEC uses a Gaussian Mixture Model to identify peaks in reciprocal space with distinct temperature dependencies. In order to distinguish the functional form  of the intensity-temperature trajectory rather than their magnitudes, the intensity of each momenta $\vec{q}$ is re-scaled (z-scored) as $\tilde{I}_{\vec{q}}(T)=\left(I_{\vec{q}}(T)-\mu_{\vec{q}}\right)/\sigma_{\vec{q}}$ where $\mu_{\vec{q}}$ is the mean over temperature $T$, and $\sigma_{\vec{q}}$ is the standard deviation in $T$ \cite{Venderley2022Proc.Natl.Acad.Sci.}.

By applying X-TEC on the CsV$_3$Sb$_5$ data, $\sim 15,000$ peaks were identified with three distinct temperature trajectories color-coded as yellow, green and blue clusters. These clusters exhibit an onset behavior around the CDW transition temperature as shown in Figs.~\ref{fig4:xtec_clustering}(a) and (b). By looking at the locations of these clusters in reciprocal space (Fig.~\ref{fig4:xtec_clustering}(c)), we find that all three clusters have the same in-plane momenta at $(H,K)\equiv (0.5,0.5)$, but differ in their out-of-plane momentum $L$. While the green clusters are located at $L=1/4$ (Fig.~\ref{fig4:xtec_clustering}(e)) corresponding to the $2\times2\times4$ structure, the blue clusters derive from the $L=1/2$ peaks (Fig.~\ref{fig4:xtec_clustering}(f)) of the $2\times2\times2$ structure, and the yellow clusters to an \textit{additional} class of $L=0$ peaks (Fig.~\ref{fig4:xtec_clustering} (d)) representing a $2\times2\times1$ structure. From their intensity-temperature trajectories (Fig.~\ref{fig4:xtec_clustering}(b)), we find that the onset of CDW proceeds in two stages. First, the $2\times2\times1$ and $2\times2\times2$ cells appear below temperature $\approx 95$~K, followed by the onset of a $2\times2\times4$ state at $\approx 85$~K. Remarkably, with the onset of the $2\times2\times4$ peaks, the $2\times2\times1$ peaks are strongly suppressed and vanish below 80~K, and, additionally, the $2\times2\times2$ peaks initially decrease with the onset of $2\times2\times4$ order and then recover their intensity at lower temperatures. This suggests an exchange of scattering weight between order parameters.

\begin{table*}[]
	\setlength{\tabcolsep}{8pt}
	\centering
	\caption{Parameters of structural models for the room temperature and the low-temperature structures of KV$_3$Sb$_5$, RbV$_3$Sb$_5$, and CsV$_3$Sb$_5$. All .cif files from the refinements can be found in the supplemental information \cite{ESI}.}
	\begin{tabular}{lcccccccccc}
		\hline
		& \multicolumn{3}{c}{KV$_3$Sb$_5$} & \multicolumn{3}{c}{RbV$_3$Sb$_5$} & \multicolumn{4}{c}{CsV$_3$Sb$_5$} \\
		\hline
		Wavelength ($\rm \AA$) & \multicolumn{10}{c}{0.41328}\\
		
		Temperature (K) & 290 & \multicolumn{2}{c}{10} & 290 & \multicolumn{2}{c}{10} & 290 & \multicolumn{2}{c}{90} & 11 \\
		
		Space group & $P6/mmm$ & \multicolumn{2}{c}{$Fmmm$} & $P6/mmm$ & \multicolumn{2}{c}{$Fmmm$} & $P6/mmm$ & \multicolumn{2}{c}{$Fmmm$} & $Cmmm$\\
		
		Type & & \multicolumn{2}{c}{$2\times2\times2$} & & \multicolumn{2}{c}{$2\times2\times2$} & &  \multicolumn{2}{c}{$2\times2\times2$} & $2\times2\times4$\\
		
		Distortion & & TrH & SoD & & TrH & SoD & & TrH & SoD & TrH/SoD \\
		
		\hline
		\hline
		
		$N_\mathrm{measured}$ & 2751 & \multicolumn{2}{c}{17014} & 2239 & \multicolumn{2}{c}{16168} & 2888 & \multicolumn{2}{c}{6788} & 24048\\
		
		$N_\mathrm{independent}$ & 232  & \multicolumn{2}{c}{2401} & 201 & \multicolumn{2}{c}{2444} & 238 & \multicolumn{2}{c}{2159} & 5002\\
		
		$N_\mathrm{sig.}$ [$I>2\sigma(I)$] & 231  & \multicolumn{2}{c}{1617} & 201 & \multicolumn{2}{c}{1906} & 238 & \multicolumn{2}{c}{641} & 3098\\
		
		$a$ ($\rm \AA$) & 5.4260(9) & \multicolumn{2}{c}{10.9576(9)} & 5.4941(6) & \multicolumn{2}{c}{11.0065(9)} & 5.5236(6) & \multicolumn{2}{c}{11.0275(9)} & 11.005(2)\\
		
		$b$ ($\rm \AA$) & 5.4260(9) & \multicolumn{2}{c}{18.9813(16)} & 5.4941(6) & \multicolumn{2}{c}{19.0575(16)} & 5.5236(6) & \multicolumn{2}{c}{19.1082(16)} & 19.046(4)\\
		
		$c$ ($\rm \AA$) & 8.845(2) & \multicolumn{2}{c}{17.8552(15)} & 9.1071(14) & \multicolumn{2}{c}{18.1010(15)} & 9.3623(15) & \multicolumn{2}{c}{18.6213(15)} & 37.176(7)\\
		
		$\rho$ (g/cm$^3$) & 5.90 & \multicolumn{2}{c}{5.73} & 5.91 & \multicolumn{2}{c}{5.93} & 6.00 & \multicolumn{2}{c}{6.06} & 7.22\\
		
		$R_1 > 4 \sigma$ (\%) & 8.62 & 6.22 & 6.31 & 10.42 & 6.40 & 7.41 & 5.91 & 7.71 & 7.73 & 10.0\\
		
		$R_1$ all (\%) & 8.63 & 6.44 & 6.56 &  10.42 & 6.74 & 7.82 & 5.91 & 9.61 & 9.62 & 11.55\\
		
		Twin fraction 1 & - & 0.35 & 0.36 & - & 0.32 & 0.32 & - & 0.31 & 0.30 & 0.31 \\
		
		Twin fraction 2 & - & 0.33 & 0.33 & - & 0.33 & 0.33 & - & 0.36 & 0.36 & 0.34 \\
		
		Twin fraction 3 & - & 0.32 & 0.31 & - & 0.34 & 0.34 & - & 0.33 & 0.34 & 0.35 \\
		\hline
	\end{tabular}
	\label{tab:refin}
\end{table*}

The 2$\times$2$\times$4 superstructure persists down to temperatures of 11~K. We note here that the in-plane and out-of-plane correlation lengths of the CDW superlattice peaks here are identical to the primary structural peaks.  This indicates that the  2$\times$2$\times$4 CDW state is long-range with correlation lengths constrained only by the sample's underlying crystallinity.  Notably, our diffraction data using this cooling/warming profile do not resolve anomalies at $T=60$~K as seen in magnetotransport experiments \cite{chen2021roton,xiang2021twofold} or a crossover to a 2$\times$2$\times$2 state similar to Ref.~\cite{Stahl22}. This difference potentially arises from subtle effects attributed to the different cooling profiles during measurement and potentially from sample dependent effects imparted during the crystal growth process.

We now discuss the structural refinement in the 2$\times$2$\times$2 regime at $T=90$~K and the 2$\times$2$\times$4 superstructure at lower temperatures. Using the same twinning approach described earlier, the 2$\times$2$\times$2 state at 90 K is best fit in the $Fmmm$ group rendering either a staggered SoD or TrH distortion (CIFs provided in the supplemental material \cite{ESI}).  This is consistent with the \textit{low temperature} state recently reported by Stahl et al. \cite{Stahl22} where only a 2$\times$2$\times$2 superstructure was observed. Upon cooling into the 2$\times$2$\times$4 state, the data at 10 K are instead best refined in the lower symmetry, base-centered $Cmmm$ group.  Fig.~\ref{fig5:cvs_ref} shows the resulting model of the superstructure consisting of vanadium layers with both SoD and TrH deformations. The three TrH layers are staggered along the $c$-axis by $0.5a$ (Fig.~\ref{fig5:cvs_ref}~(d),(e),(f)) and are then capped by a SoD layer (Fig.~\ref{fig5:cvs_ref}~(c)). The model assumes a uniform phase at this temperature and represents the measured diffraction data well, especially considering the large number of significant reflections (3098 shown in Tab.~\ref{tab:refin}). This mixed distortion-type result is consistent with previous refinements in a $P\overline{3}$ average structure containing both SoD and TrH layers \cite{Ortiz21superconductivity} as well as recent angle-resolved photoemission data \cite{CominCVSRVSKVS, MingShiCVS}.  Mixed reports from NMR measurements \cite{Kun, Luo2022} provide a picture interpreted as either SoD or TrH distortion types, and it is unclear whether a mixed distortion-type structure can also be modeled via the data.  Moreover, our X-TEC analysis indicates that a 2$\times$2$\times$2 phase competes with and potentially coexists with the 2$\times$2$\times$4 phase at lowest temperatures and that the stacking sequence of the TrH- and SoD-like layers may be influenced by this coexistence.

\subsection{Pulsed magnetic field experiment}

Scanning tunneling microscopy measurements have reported that a magnetic field couples to the apparent Fourier weight attributed to CDW order at the surface in $A$V$_3$Sb$_5$ compounds \cite{Jiang21}, suggesting the presence of a chiral CDW state that breaks time reversal symmetry.  As means of testing whether a magnetic field couples to the real component of the CDW order parameter in the bulk, the magnetic field dependence of the superlattice reflections was investigated.  A pulsed magnetic field of $H_\mathrm{pulse} = 28$~T was used, and the geometry of the inner bore of the magnet coil allows for the measurement of the \textbf{Q}$=(0.5,0.5,0.25)$ superlattice reflection in transmission geometry with the field applied parallel to the $c$-axis.  Scattering data were collected at $T=11$~K and at $T = 82.5$~K after cooling the sample into the CDW state under zero field conditions. Magnetic field pulses were applied that last for approximately 10~ms (half-sine, start to finish), and where the maximum magnetic field of 28 T is reached.  Multiple pulses were used to improve measurement statistics. 

The \textbf{Q}$=(0.5,0.5,0.25)$ peak was measured at $T = 11$~K for a total of 45 pulses and at $T = 82.5$~K for a total of 61 pulses.  For illustration, Fig.~\ref{fig6:cvs_h_puls}~(a) shows the result of 1 ms of data collection during a single field pulse on the \textbf{Q}$=(0.5,0.5,0.25)$ superlattice peak on a pixel array. Fig.~\ref{fig6:cvs_h_puls}~(b) then shows the same peak after a cumulative sum of 61 pulses. Each time bin corresponds to a field value shown in Fig.~\ref{fig6:cvs_h_puls}~(c).  

The integrated intensities of the \textbf{Q}$=(0.5,0.5,0.25)$ peak as a function of time were summed over the number of individual pulses to increase the signal-to-noise ratio with the results plotted in Fig.~\ref{fig6:cvs_h_puls}~(d). No significant change of the integrated intensity outside of the experimental error could be observed at either of the two temperatures explored. The pre-pulse fluctuations and sharp post-pulse decrease in the integrated intensity are attributed to mechanical vibrations. Within our experimental resolution, we do not observe any evidence of magnetic-field dependence of the superlattice peak intensity, consistent with a recent report \cite{Stahl22} and signifying that the real component of the CDW order parameter is unchanged. 

\section{Discussion and Conclusions}

Tab.~\ref{tab:refin} summarizes the parameters of the crystallographic refinements carried out on KV$_3$Sb$_5$, RbV$_3$Sb$_5$, and CsV$_3$Sb$_5$ crystals at key temperatures discussed in the preceding sections. All refined models represent the collected data well and the $R$-values are low considering the high number of significant reflections collected in synchrotron experiments.  The CDW distortion results in an average V-V bond length that is nearly the same in all three AV$_3$Sb$_5$ variants with (V-V)$_{avg}$=2.7022$\pm0.022$ in A = K, (V-V)$_{avg}$=2.6992$\pm0.028$ in A = Rb, and (V-V)$_{avg}$=2.6979$\pm0.036$ in A = Cs.  The standard deviations in these V-V distances highlight the fact that the assignments of TrH or SoD distortions are only effective classifications and that the V-V bond lengths have substantial variance within a given layer.

The increase in the standard deviation of distorted bond lengths about a fixed average value likely indicates a progressively increasing instability of the TrH state as the A-site changes from K to Rb to Cs.  Notably, the Cs system has the largest room temperature V-V distance, rendering its low-temperature reconstruction of the kagome network the most dramatic of the series.  This is likely reflected in the larger heat capacity anomaly in the Cs system relative to the other two as well as its accommodation of a mixed TrH and SoD distortion.

The mixed SoD and TrH state refined in the CDW state of CsV$_3$Sb$_5$ should also be viewed through the lens of the experimental parameters used in this study.  Slower cooling or faster quenching may modify the resulting structure in the CDW state given the seeming competition with the $2\times 2\times 2$ TrH state and a newly observed $2\times 2\times 1$ state that both appear just below $T_{CDW}$.  It is possible that materials with different levels of strain frozen during the growth process or dilute impurity content may modify this behavior as well.  Subtle disorder along the interlayer direction in the form of variable crystallinity between samples does not seem to play a dominant role as we have screened many crystals with variable quality and they all exhibit the same $2\times 2\times 4$ low-temperature superlattice using this same cooling profile.

In summary, we studied the structural deformations related to the CDW transitions in the systems CsV$_3$Sb$_5$, KV$_3$Sb$_5$, and RbV$_3$Sb$_5$ using synchrotron x-ray diffraction. Our  diffraction data allow for twinned models of orthorhombic superstructures in the CDW state. We find 2$\times$2$\times$2 superstructures in KV$_3$Sb$_5$ and RbV$_3$Sb$_5$ and our crystallographic refinements allow us to parameterize the distortion of vanadium-based kagome layers in a staggered TrH state. Temperature dependent diffraction data from the CsV$_3$Sb$_5$ system indicate a complex evolution of the structural superlattice. Initially, a 2$\times$2$\times$1 state appears along with a 2$\times$2$\times$2 state at temperatures below 90~K, and this is followed by the onset of a 2$\times$2$\times$4 superlattice below 85 K. Structural refinement of this larger superstructure reveals a three-layer staggered TrH deformation capped by a layer with a SoD type distortion. Scattering data from CsV$_3$Sb$_5$ measured under a 28 T pulsed magnetic field fails to resolve a change in the CDW superlattice intensity, setting bounds on the response of the real component of the CDW state to the application of a magnetic field. 

\section{Acknowledgments}
 SDW acknowledges helpful discussions with Turan Birol, Rafael Fernandes, and Binhai Yan. This work was supported by the National Science Foundation (NSF) through Enabling Quantum Leap: Convergent Accelerated Discovery Foundries for Quantum Materials Science, Engineering and Information (Q-AMASE-i): Quantum Foundry at UC Santa Barbara (DMR-1906325). The research made use of the shared facilities of the NSF Materials Research Science and Engineering Center at UC Santa Barbara (DMR- 1720256). The UC Santa Barbara MRSEC is a member of the Materials Research Facilities Network. (www.mrfn.org). This work is based upon research conducted at the Center for High Energy X-ray Sciences (CHEXS) which is supported by the National Science Foundation under award DMR-1829070.  The high-field pulsed magnet and a choke coil were installed at the Advanced Photon Source through a partnership with International Collaboration Center at the Institute for Materials Research (ICC-IMR) and Global Institute for Materials Research Tohoku (GIMRT) at Tohoku University. The development and use of X-TEC by K.M. and E.-A.K. was supported by the U.S. Department of Energy, Office of Basic Energy Sciences, Division of Materials Sciences and Engineering. KM and E-AK are supported in part by the Gordon and Betty Moore Foundation’s EPiQS Initiative, Grant GBMF10436.

\bibliography{references_avs_struc}

\end{document}